\documentclass[preprint,aps,draft,amsmath,amssymb]{revtex4}

\usepackage{graphicx}% Include figure files
\usepackage{dcolumn}% Align table columns on decimal point
\usepackage{bm}% bold math

\newcommand{\resp}{{\it resp.\ }}
\newcommand{\calh}{\mathcal{H}}
\newcommand{\ket}[1]{| #1 \rangle}
\newcommand{\bra}[1]{\langle #1 |}
\newcommand{\tr}{\text{tr}}

\begin{document}

\title{Non-Bilocal Measurement via Entangled State}% Force line breaks with \\

\author{Eran Shmaya}
\affiliation{School of Mathematical Sciences, Tel Aviv
University}\email{gawain@post.tau.ac.il}
 %\altaffiliation{School of Mathematics, Tel Aviv University}\email{gawain@post.tau.ac.il}%Lines break automatically or can be forced with \\
%\affiliation{%
%\affiliation{

\date{\today}% It is always \today, today,

\begin{abstract}
Two observers, who share a pair of particles in an entangled mixed
state, can use it to perform some non-bilocal measurement over
another bipartite system. In particular, one can construct a
specific game played by the observers against a coordinator, in
which they can score better than a pair of observers who only share
a classical communication channel. The existence of such a game is
an operational implication of an entanglement witness.
\end{abstract}

%                             % Classification Scheme.
\maketitle
\section{Introduction}
The relationship between entanglement and non-locality of quantum
systems has been the subject of extensive research. The most
celebrated manifestation of the non-local aspect of entanglement
is Bell's theorem~\cite{bell}, that correlations of outcomes of
measurements over a pair of particles at singlet state cannot be
squared with a local hidden variable model. This theorem was later
extended to every pure entangled state~\cite{gissinperes}. The
case of mixed states is more challenging. Werner~(\cite{werner})
constructed an entangled bipartite state that admits a local
hidden variables model which reproduces all the statistical
correlations of von-Neumann (ideal) measurements over the
subsystems. In particular, the correlations of outcomes of local
ideal measurements on a pair of particles at Werner's state do not
violate any Bell inequality. But it was later discovered that
Werner's states manifest some other non-local
aspects~\cite{popescu_tlprt,popescu_density,peres_collective}. The
question therefore arises, does every entangled state manifest
some aspect of non-locality?

The concept of entanglement is easily generalized from pure states
to mixed states. A nonnegative operator $F$ over a tensor product
of Hilbert spaces is called \emph{separable} if it can be written
in the form $\Sigma_i K_i\otimes K_i'$, where $K_i,K_i'$ are
nonnegative operators. A mixed state of a bipartite quantum system
is called \emph{entangled} if the corresponding density operator
is non-separable. These are well defined mathematical concepts,
which are somehow related to the more vague physical concept of
non-locality. As mentioned above, several aspects of non-locality
have been suggested in the literature. The purpose of this paper
is to present a new facet of non-locality, that is manifested by
\emph{any} entangled state: Observers who share a pair of
particles in that state can use it to perform non-bilocal
measurement over another pair.

A measurement (or a POVM measurement) is represented by a
$k$-tuple of nonnegative operators $(F_1,\dots,F_k)$ such that
$F_1+\dots+F_k=I$. If the state of a system is represented by the
density operator $W$ and the POVM measurement $(F_1,\dots,F_k)$ is
performed over the system, the outcome is $i$ with probability
$\tr(W\cdot F_i)$. Particularly important for this paper is the
case $k=2$, i.e. measurements with two possible outcomes, `yes'
and `no'. We call such measurements \emph{yes-no measurements}. A
yes-no measurement is given by an operator $F$ such that $0\leq
F\leq I$. If the state of a system is $W$ and the yes-no
measurement $F$ is applied, the measurement's outcome is `yes'
with probability $\tr(W\cdot F)$, and `no' with probability
$1-\tr(W\cdot F)$. Yes-no measurements are called effects
in~\cite{kraus}. A POVM measurement $(F_1,\dots,F_k)$ is called
\emph{local} if it can be carried out by Alice or Bob. This means
that either $F_i=F_i^{(A)}\otimes I$ for each $i$ (in which case,
to perform the measurement Alice alone has to perform the
measurement $(F_1^{(A)},\dots,F_k^{(A)})$ on her particle), or
that $F_i=I\otimes F_i^{(B)}$ for each $i$. A POVM measurement is
called \emph{bilocal}~(\cite{nlclt_entgl}) if it can be performed
by a sequence of local measurements and classical communication.
Note~(\cite{nlclt_entgl}) that the operator $F$ corresponding to a
yes-no bilocal measurement is necessarily separable.

In order to get some intuition about how mixed entangled states can
be used to perform non-bilocal measurements, we first consider two
examples. Assume that all particles have spin $\frac{1}{2}$ and that
Alice and Bob share a pair of particles in a singlet state, that is
given by the density operator
$\rho_{\text{singlet}}=\frac{1}{2}(\ket{01}\bra{01}-\ket{01}\bra{10}-\ket{10}\bra{01}+\ket{10}\bra{10})$.
If they are now introduced to another pair of particles at unknown
state $W$, they can use the singlet pair $\rho_{\text{singlet}}$ to
teleport~\cite{teleportation} Alice's part of $W$ to Bob. Bob now
holds a pair of particles at state $W$, to which he can apply any
yes-no measurement. Thus, using the singlet pair, Alice and Bob are
able to perform non-bilocal measurements over the new pair. In
particular they can deduce more information about the unknown state
$W$ than can a pair of observers who can only communicate
classically.

Consider another example. Suppose that Alice and Bob share a pair
of particles at Werner's state, which is given by the density
operator $\rho_W=\frac{1}{2}\rho_{\text{singlet}}+\frac{1}{8}I$,
where $I$ is the identity operator. Werner~(\cite{werner}) showed
that, even though this state is entangled, there exists a local
hidden variables model that reproduces the correlations of all
ideal local measurements that can be performed on it. Still, as
was shown by Popescu~(\cite{popescu_tlprt}), the non-local aspect
of Werner's state is revealed when one tries to use it in the
teleportation scheme instead of the singlet. This yields
teleportation with better fidelity than the maximal fidelity that
can be achieved by using only classical communication. We now show
how this imperfect teleportation can be used to perform some
non-bilocal measurement over a pair of particles at state $W$.
Assume that Alice and Bob try to transfer Alice's part of $W$ to
Bob using the teleportation scheme with Werner's state $\rho_W$.
Note that Werner's state can be seen as a mixture of the singlet
$\rho_{\text{singlet}}$ with a completely random state
$\frac{1}{4}I$. Thus, with probability $0.5$ the teleportation
succeeds and Bob holds a pair of particles in state $W$. With
probability $0.5$ the teleportation fails, transferring the
completely random spin-$\frac{1}{2}$ particle at state
$\frac{1}{2}I$ to Bob. Thus after this process Bob holds a pair of
particles at state $\frac{1}{2}W+\frac{1}{4}I\otimes \tr_A(W)$,
where $tr_A(W)$ is the partial trace over subsystem $A$ of $W$
(which represents the state of Bob's part of $W$ before the
measurement). Suppose now that Bob performs the yes-no measurement
given by the operator $\rho_{\text{singlet}}$ on this pair. The
probability to receive outcome `yes' is given by
\[\tr\left(\left(\frac{1}{2}W+\frac{1}{4}I\otimes
\tr_A(W)\right)\cdot\rho_{\text{singlet}}\right)=
\tr(W\cdot\rho_W).\] Thus using local measurements and classical
teleportation, Alice and Bob simulated the yes-no measurement given
by the operator $\rho_W$. Since $\rho_W$ is non-separable, this is a
non-bilocal measurement.

Thus, the non-locality of Werner's state is revealed by the fact
that observers can use it to perform a non-bilocal measurement.
The purpose of this paper is to show that \emph{every} entangled
state $\rho$ manifests this aspect of non-locality: A pair of
observers who share this state can use it to perform some
non-bilocal yes-no measurement. In section (\ref{game}) the
possibility of performing non-bilocal yes-no measurement using an
entangled state is given a game theoretic interpretation: We
consider game played by a pair of players, Alice and Bob, against
a game coordinator, in which Alice and Bob have to guess the state
of a bipartite system prepared by the coordinator. It is shown
that if Alice and Bob share an entangled state they gain an
advantageous guessing strategy by using the non-bilocal
measurements.

It is interesting to compare the result of this paper with another
aspect of non-locality, namely distillation. An entangled state
$\rho$ is called \emph{distillable}, if it is possible to create
,with high probability, a singlet state from a large set of copies
of $\rho$ using only local operations and classical communication.
It is known (\cite{bennet-dist,hor-mixed-dist}) that every pure
entangled state is distillable, but there exist mixed states which
cannot be distilled. These states are sometimes called \emph{bound
entangled states}. The fact that for every entangled state $\rho$
there exists a state-guessing game in which sharing $\rho$ is
advantageous shows that even bound entangled states are still
useful in certain situations.

The link between the non-bilocal measurement presented in
Section~\ref{protocol} and the state-guessing game presented in
Section~\ref{game} is an entanglement witness. Entanglement
witness can be viewed geometrically as a hyperplane that separates
an entangled state from the convex set of separable states. It is
known (\cite{terhal}) that every entangled state $\rho$ admits an
entanglement witness and (\cite{geometric-bell}) that the distance
between $\rho$ and the set of separable states in the Euclidian
space of Hermitian operators equals the maximal violation of the
corresponding ``generalized Bell inequality'' (see also
\cite{reflections} for relationship between entanglement witnesses
and distillation and \cite{key-distribution} for the use of
certain entanglement witnesses to prove the presence of
entanglement in order to establish a secure key distribution.) The
fact that every entanglement witness gives rise to a specific
game, in which the players benefit from sharing $\rho$ is an
operational implication of the entanglement witness. Thus, this
paper shows that the existence of an entanglement witness is not
only necessary for a state to be entangled, but is also sufficient
for the state to reveal non-locality.

\section{Scheme for Non-Bilocal Measurement}\label{protocol}
In this section we describe a scheme for performing a non-bilocal measurement
using a pre-prepared entangled pair $\rho$.

Let $\rho$ be a non-separable density matrix over
$\calh_A\otimes\calh_B$. Consider a pair of particles at state
$\rho$ and assume that Alice has access to the particle that lives
in $\calh_A$ and Bob has access to the particle that lives in
$\calh_B$. Assume now that Alice and Bob are introduced to another
pair of particles at the unknown state represented by the density
matrix $W$ over $\calh_A'\otimes\calh_B'$, such that
$\dim(\calh'_A)=\dim(\calh_A)=n$ and
$\dim(\calh_B')=\dim(\calh_B)=m$. Thus, the joint state of the $4$
particles is represented by the density matrix $\rho\otimes W$ over
$\calh_A\otimes\calh_B\otimes\calh_A'\otimes\calh_B'$. Alice has
access to the subsystem $\calh_A\otimes\calh_A'$ and Bob has access
to the subsystem $\calh_B\otimes\calh_B'$.

Let $\{\ket{i}\},\{\ket{i'}\},\{\ket{\mu}\},\{\ket{\mu'}\}$ be
orthogonal bases for $\calh_A,\calh_A',\calh_B,\calh_B'$ \resp Note
that Latin indices correspond to the particles held by Alice and
Greek indices correspond to the particles held by Bob. Let
$\ket{\phi_A}=\frac{1}{\sqrt{n}}\Sigma_i\ket{i}\otimes\ket{i'}$ and
$\ket{\phi_B}=\frac{1}{\sqrt{m}}\Sigma_\mu\ket{\mu}\otimes\ket{\mu'}$.
Assume that Alice and Bob perform the yes-no measurement
$\ket{\phi_A}\bra{\phi_A}\otimes\ket{\phi_B}\bra{\phi_B}$ on the
$4$-particle system
$\calh_A\otimes\calh_A'\otimes\calh_B\otimes\calh_B'$. Note that
this can be done by local measurements and classical communication:
Alice performs the yes-no measurement $\ket{\phi_A}\bra{\phi_A}$
over $\calh_A\otimes\calh_A'$, Bob performs the yes-no measurement
$\ket{\phi_B}\bra{\phi_B}$ over $\calh_B\otimes\calh_B'$, and the
outcome of the measurement is given by the logical conjunction of
the local outcomes received by Alice and Bob (thus, classical
communication is needed to establish the outcome of the measurement
from the outcome of the local measurements.)

One can verify that, for every density matrix $W$ over
$\calh_A\otimes\calh_B$,
\[\tr\left((\ket{\phi_A}\bra{\phi_A}\otimes
\ket{\phi_B}\bra{\phi_B})\cdot(\rho\otimes
W)\right)=\frac{1}{nm}\Sigma_{i,j,\mu,\nu}\bra{i\mu}
\rho\ket{j\nu}\bra{i'\mu'}W\ket{j'\nu'}=\frac{1}{nm}\tr(W\cdot\rho^t),\]
where $\rho^t$ is the transpose of $\rho$ w.r.t the basis
$\{\ket{i\mu}\}_{i,\mu}$ of $\calh_A\otimes\calh_B$. Thus, this
scheme effectively performs the yes-no measurement
$\frac{1}{nm}\rho^t$ over $W$. But since $\rho$ is a non-separable
matrix, it follows that $\frac{1}{nm}\rho^t$ is also non-separable.
Thus using this scheme, Alice and Bob perform a non-separable, and,
in particular non-bilocal measurement over the state $W$.

\section{A State-Guessing Game}\label{game}
In this section we try to shed some light on the implications of the
non-bilocal measurement constructed above. To do so, we describe a
specific \emph{game} that Alice and Bob play against a game
coordinator, in which they can use the non-bilocal yes-no
measurement $\frac{1}{nm}\rho^t$ to score better than a pair of
observers who can only communicate classically. The discussion
follows standard game-theoretic arguments.

Let $H$ be an \emph{entanglement witness} (\cite{terhal}), i.e an
Hermitian operator such that $\tr(H\cdot \rho) < 0$ but
$\tr(H\cdot D) \geq 0$ for every separable $D$. The existence of
such an operator $H$ follows from the inseparability of $\rho$ and
the separation theorem for convex cones~(\cite{convex}). Let $H^t$
be the transpose of $H$ w.r.t the basis $\{\ket{i\mu}\}_{i,\mu}$
of $\calh_A\otimes\calh_B$. We can assume that $H^t=\beta
W^2-\alpha W^1$ where $W^1$ and $W^2$ are density operators, and
$\beta,\alpha\geq 0$. Since $\tr(H^t)=\tr(H)\geq 0$, it follows
that $\beta\geq \alpha$.

Suppose that Alice and Bob are engaged in the following game: At the
beginning of the game, a pair of particles is prepared by the game
coordinator at state $W^1$ or $W^2$ with probabilities
$\frac{\alpha}{\alpha+\beta},\frac{\beta}{\alpha+\beta}$ \resp The
first particle is given to Alice and the second to Bob. Alice and
Bob, who share a classical communication channel, know the
parameters of the game (i.e $W^1,W^2,\alpha,\beta$,) and their goal
is to guess which state was actually chosen. They receive payoff
$+1$ for a correct guess and $-1$ for an incorrect guess.

Every strategy that Alice and Bob can apply in the game corresponds
to some yes-no measurement $F$ on the pair of particles: If the
outcome of the measurement is `yes' they guess that the state was
$W^1$, if the outcome is `no' they guess that the state was $W^2$.
Their expected payoff is thus given by
\begin{multline*}\frac{\alpha}{\alpha+\beta}\left (\tr(W^1\cdot
F))-\tr(W^1\cdot(I-F)\right
)\\+\frac{\beta}{\alpha+\beta}\left(\tr(W^2\cdot(I-F))-\tr(W^2
\cdot
F)\right)=\frac{\beta-\alpha}{\alpha+\beta}-\frac{2}{\alpha+\beta}\tr(H^t\cdot
F).\end{multline*}

If Alice and Bob can only perform local measurements and
communicate classically, the yes-no measurement $F$ they employ is
necessarily a separable operator, and their expected payoff is
therefore no greater than $\frac{\beta-\alpha}{\alpha+\beta}$. If,
on the other hand, Alice and Bob share a bipartite system at state
$\rho$, they can implement the scheme described in
section~\ref{protocol} and thus achieve a payoff
$\frac{\beta-\alpha}{\alpha+\beta}-\frac{2}{nm(\alpha+\beta)}\tr(H^t\cdot\rho^t)$.
Since $\tr(H^t\cdot\rho^t)=\tr(H\cdot\rho)<0$ this is strictly
greater than the payoff they can achieve without this system.

%proofs
%\[
%\frac{\alpha}{\alpha+\beta} \{ [\tr(W^1F)]-\tr[W^1(1-F)] \} +
%\]
%$\tr(H^t\rho^t)$ $\tr(H\rho)$

\begin{acknowledgments}
This paper is written during my Ph.d. study in Tel Aviv
University. I am grateful to my supervisor Prof. Ehud Lehrer for
the time and ideas he shares with me.
\end{acknowledgments}

\end{document}